\documentclass[twocolumn,aps,pre,floatfix,nofootinbib]{revtex4-2}
\usepackage{psfrag,epsfig,amsfonts,amssymb,amsmath,wasysym,bm}
\usepackage{dcolumn}
\usepackage{bbold}
\usepackage{color}
\usepackage{soul}
\usepackage[normalem]{ulem}
\usepackage{tabularx}
\usepackage{tikz}

\usepackage[inline]{enumitem} 

\newcommand{\RR}{{\mathbb R}}

\newcommand{\tr}{\mbox{Tr}}
\newcommand{\hr}{{\cal H}}

\providecommand{\opnorm}[1]{\|#1\|_{\!\!\; {\rm op}}}

\newcommand{\lmat}{\left( \begin{matrix}}	
\newcommand{\rmat}{\end{matrix} \right)}	

\begin{document}

\title{Similarities and differences of typicality in quantum and classical systems}

\author{Peter Reimann}
\affiliation{Faculty of Physics, 
Bielefeld University, 
33615 Bielefeld, Germany}

\author{Nicol\'as Nessi}
\affiliation{1 IFLP CONICET,
Diagonal 113 y 64, La Plata, Buenos Aires, Argentina}

\date{\today}

\begin{abstract}
Typicality is a well-established and very general property of quantum 
many-body systems, 
referring to the phenomenon that the expectation 
values of any given observable are 
practically indistinguishable
for the overwhelming 
majority of all pure states (normalized vectors) 
in a sufficiently high-dimensional Hilbert 
(sub-)space.
Here, we provide very simple and general arguments that analogous 
typicality properties of pure states (phase space points) in 
classical many-body systems are still 
expected to hold true
for macroscopic 
observables, 
but not any more for microscopic (few-body) observables.
\end{abstract}

\maketitle

%
\section{Introduction}
\label{s1}

Essentially, 
quantum
typicality refers to the following quite remarkable 
feature
of any sufficiently high-dimensional Hilbert space $\hr$:
Consider an arbitrary subspace $\hr_{\! D}\subseteq\hr$ of large but finite dimension $D$,
and let us randomly sample
pure states from $\hr_{\! D}$, i.e., vectors $|\psi\rangle$ which are normalized 
but otherwise unbiased (uniformly distributed \cite{f1}).
Then the expectation values 
$\langle \psi | A |\psi\rangle$ of any given observable 
(Hermitian operator) $A$ will assume with very high probability
very similar values.

To the best of our knowledge, this so-called typicality or 
concentration of measure phenomenon 
has been 
originally discovered in Ref. \cite{llo88},
and has later been independently rediscovered 
several times
\cite{gol06,pop06,llo06,gem04,sug07,bal08}.
A more detailed 
review of the pertinent literature
and of various modifications and generalizations 
can be found for instance in 
Ref. \cite{rei20}.

A very natural next question is whether or not analogous 
typicality 
features 
may also be encountered in classical 
systems.
For a few special model classes this questions has been 
addressed for instance in Refs.~\cite{yos15,nes25a,nes25b}.
Yet statements of comparable rigor, simplicity, and 
generality as in the quantum case 
are not known until now.
The main message of our present paper is that
such statements
can indeed be obtained relatively easily
provided one is willing to 
take for granted
some
physically very convincing but not yet rigorously
proven 
assumptions.

%
\section{Typicality in quantum systems}
\label{s2}

For the sake of completeness, we first provide a somewhat more 
precise formulation of the quantum typicality property from Sec.~\ref{s1}.
In a second step, we point out some rather simple but little known
implications.

Returning to the quantum setting from the first paragraph 
of Sec.~\ref{s1}, typicality may be more precisely 
characterized as follows.
If we denote by $\bar A$ 
the average over all those randomly sampled expectation values $\langle \psi | A |\psi\rangle$, 
then the probability that the deviation of $\langle \psi | A |\psi\rangle$ 
from $\bar A$
exceeds an arbitrary but fixed threshold 
value $\delta>0$ can be upper bounded as
 \begin{eqnarray}
 \mbox{Prob}\!\left( | \langle \psi | A |\psi\rangle - \bar A | >\delta \right)
 \leq 
 \frac{\opnorm{A}}{\sqrt{D\delta}}
 \ ,
 \label{n1}
 \end{eqnarray}
where $\opnorm{A}$ denotes the operator
norm of $A$ (largest eigenvalue in modulus)
 \cite{f0}.
Given that $A$ models some experimental measurement device, the eigenvalues of $A$
correspond to the possible measurement outcomes, hence $\opnorm{A}$ will be
bounded and usually can be set equal to unity without much loss of generality
\cite{f4}.
Choosing for $\delta$ the resolution limit of the given measurement device, 
and assuming that $D$ is much larger than $1/\delta$, 
it follows that the expectation values 
$\langle \psi | A |\psi\rangle$ will be experimentally indistinguishable from 
$\bar A$ (and thus from each other) for the overwhelming majority 
of all normalized vectors (pure states) $|\psi\rangle\in\hr_{\! D}$.
In brief, $\langle \psi | A |\psi\rangle$ is typically very close to $\bar A$.

Moreover, the averaged expectation value $\bar A$ 
in (\ref{n1})
can 
be readily
rewritten as 
\begin{eqnarray}
\bar A = \tr\{\bar \rho A\}
\ ,
\label{n2}
\end{eqnarray} 
where $\bar \rho$ is the statistical operator 
which arises
by averaging $|\psi\rangle\langle\psi|$ over all the
uniformly distributed random
pure states $|\psi\rangle$. 
It follows that $\bar\rho$ is independent of $A$,
is generically a mixed state ($\tr\{\bar\rho^2\}<1$),
and can be imitated very well 
(as far as expectation values are concerned) by
most pure states $|\psi\rangle\in\hr_{\! D}$.


Examples of foremost physical interest
are quantum
many-body 
system  with $N\gg 1$ degrees of freedom, 
Hamiltonian $H$, and concomitant eigenvectors  
$|n\rangle$ and eigenvalues $E_n$.
As usual in the context of the microcanonical formalism,
we furthermore focus on some microcanonical energy 
window 
\begin{eqnarray}
I:=[E,E+\Delta]
\label{n3}
\end{eqnarray}
 of macroscopically small but 
microscopically large width $\Delta$, meaning that $\Delta$ is experimentally 
unresolvably small while the number of states $|n\rangle$
with the property $E_n\in I$ is very large 
(exponentially growing with $N$). 
As our $\hr_{\! D}$ we choose the 
corresponding
subspace spanned 
by all those $|n\rangle$ with $E_n\in I$, which is often
denoted as the microcanonical energy shell in this context.
Hence, the above introduced statistical operator $\bar \rho$ can be
identified with
the microcanonical ensemble  
\begin{eqnarray}
\rho_{\rm mc}:=P/\tr\{P\}
\ ,
\label{n4}
\end{eqnarray} 
where $P$ is the projector onto $\hr_{\! D}$, 
and where
$\tr\{P\}$ can be identified with the dimensionality $D$ 
of the energy shell $\hr_{\! D}$.
Usually, the latter
will thus be
exponentially large in the system's degrees of freedom $N$.
Moreover, the 
averaged expectation value
$\bar A$ 
in (\ref{n1})
can be identified 
by means of (\ref{n2})
with the microcanonical 
expectation value
\begin{eqnarray}
A_{\rm mc}:=\tr\{\rho_{\rm mc}A\}
\ .
\label{1}
\end{eqnarray}

%
\subsection{Implications and remarks}
\label{s21}

In summary, typicality thus implies that for a randomly sampled 
$|\psi\rangle\in\hr_{\! D}$ 
the expectation value $\langle \psi | A |\psi\rangle$ is practically
indistinguishable from the thermal equilibrium value $A_{\rm mc}$ 
with overwhelming probability
(exceptions are exponentially rare in $N$).

Importantly, all these findings are entirely independent of 
the detailed properties of the Hamiltonian $H$
which governs the time-evolution of some given model system.
For instance, $H$ may or may not be integrable, exhibit many-body
localization (MBL), satisfy the eigenstate thermalization 
hypothesis (ETH), and so on.
On the other hand, this means that typicality does not admit
any conclusions 
regarding the time-evolution of some given model system,
in particular its long-time properties such as equilibration 
and thermalization.
Finally, also the question of whether the state 
of some given system can be decently modelled by 
such a randomly sampled vector $|\psi\rangle$
may often be a very subtle issue in itself.
For instance, this is obviously impossible
for the initial state of a system which is known to 
be out of equilibrium.

While all this is by now widely known, we finally mention
some immediate consequences which seem to be less known, but 
are of notable physical interest in themselves.
Our first observation is that if $A$ 
models some experimental measurement device, the same applies 
to $A^2$ (one simply has to square the outcome of every 
measurement of $A$),
and analogously for $A-A_{\rm mc}$
and hence for $(A-A_{\rm mc})^2$.
Typicality thus implies that the approximation
\begin{eqnarray}
\langle \psi| (A-A_{\rm mc})^2|\psi\rangle = \tr\{\rho_{\rm mc} (A-A_{\rm mc})^2\}
\label{2}
\end{eqnarray}
is very well satisfied for most $|\psi\rangle\in\hr_{\! D}$.
Moreover, $A_{\rm mc}$ on the left-hand side can be very
well approximated by $\langle\psi|A|\psi\rangle$ for most
$|\psi\rangle\in\hr_{\! D}$, as seen above.
One readily can conclude that both the latter approximation as 
well as the approximation (\ref{2}) are simultaneously satisfied 
very well for most $|\psi\rangle\in\hr_{\! D}$,
implying that the same also applies to the approximation
\begin{eqnarray}
\langle \psi| (A-\langle\psi|A|\psi\rangle)^2|\psi\rangle = \tr\{\rho_{\rm mc} (A-A_{\rm mc})^2\}
\ .
\label{3}
\end{eqnarray}
The square root of the quantity on the left-hand side is usually considered to represent
the quantum fluctuations or quantum uncertainty of $A$ 
in the pure state $|\psi\rangle$.
Indeed, it is exactly this type of expression that appears in 
Heisenberg's uncertainty principle.
On the other hand, the square root of the 
right-hand side of (\ref{3}) is commonly 
identified with the thermal fluctuations of $A$
(statistical fluctuations of $A$ in the thermal equilibrium 
ensemble $\rho_{\rm mc}$).
In other words, typicality implies that {\em thermal equilibrium fluctuations are
very well imitated by the purely quantum mechanical 
uncertainties (quantum fluctuations) for most pure state 
$|\psi\rangle$ in the energy shell.}

Particularly interesting situations arise when the thermal fluctuations
on the right-hand of (\ref{3}) are {\em not} unobservably small. 
A first example are the well-known critical fluctuations arising in the
context of phase transitions for systems at a critical point.
A second example is the position 
of a Brownian particle suspended in a liquid.
In such cases, also the quantum fluctuations on the left-hand side
of (\ref{3}) will be of macroscopic size for most pure states 
$|\psi\rangle$. In other words, they amount to so-called 
Schr\"odinger cat-states, which are generally considered 
not to be experimentally feasible.
This raises the interesting, and to our knowledge previously never 
addressed issue of how many of the typical states $|\psi\rangle$ 
are in fact unfeasible in some given model system.

%
\section{Typicality in classical systems}
\label{s3}


Similarly as in the quantum case, we 
focus on classical 
many-body systems, whose pure states are described by
phase space points $\phi$,
and whose dynamics is governed by some 
Hamilton function $H(\phi)$, while
observables are now modelled by phase-space functions $A(\phi)$.
As before, 
the quantity $I$ in Eq. (\ref{n3})
denotes some microcanonical energy 
window, while the concomitant energy shell is now defined as the 
subset of all phase space points $\phi$ with the property that 
$H(\phi)\in I$. Furthermore, the microcanonical ensemble $\rho_{\rm mc}(\phi)$ 
is defined as some suitably normalized function which is constant 
within the energy shell and zero otherwise, giving rise to
thermal (microcanonical) expectation values of the form
\begin{eqnarray}
A_{\rm mc}:=\int d\Gamma \rho_{\rm mc}(\phi) A(\phi)
\ ,
\label{4}
\end{eqnarray}
where $\int d\Gamma$ indicates the integral over the entire 
phase space, and moreover incorporates some suitable (textbook)
normalization constant,  accounting for the correct dimensionality
of the integral and the proper counting of indistinguishable particles.
Accordingly, the thermal equilibrium fluctuations of $A(\phi)$ can
be quantified by
\begin{eqnarray}
\sigma_{\! A}^2 := \int d\Gamma \rho_{\rm mc}(\phi) (A(\phi)-A_{\rm mc})^2
\ .
\label{5}
\end{eqnarray}
Overall, the similarities and differences compared to the quantum case
(previous section) are obvious and well-known.

Let us first focus on observables $A(\phi)$ which depend non-trivially 
on only a few of the very many components of $\phi$, and which
therefore may be denoted as microscopic observables.
A particularly simple example is one component of the
momentum of one particle (in cartesian coordinates).
In the latter special case, the fluctuations of the observable $A(\phi)$ 
are well-known to be governed (in very good approximation for 
sufficiently large systems) by the equipartition theorem,
implying $\sigma_{\! A}^2=mk_BT$, 
where $m$ is the mass of the 
considered particle, $k_B$ Boltzmann's constant, 
and $T$ the temperature associated with the microcanonical
ensemble $\rho_{\rm mc}$.
The salient point is that
those fluctuations are generally {\em not} negligible on the
characteristic (microscopic) scale of the considered 
observable.
It seems reasonable to expect that a qualitatively 
similar behavior will also be recovered for more
general microscopic observables $A(\phi)$:
Whenever looking at some system properties on the
microscopic scale, the statistical fluctuations 
(at thermal equilibrium) are expected to be non-negligible.
Note that while it is physically 
very reasonable to expect that this argument is 
quite generally correct, a mathematical rigorous
justification is to our knowledge not yet available.

Taking the above argument for granted, it immediately
follows that the difference $A(\phi)-A_{\rm mc}$ on the
right-hand side of (\ref{5}) cannot be negligibly small 
for most $\phi$ in the energy shell.
In other words, the system does {\em not}
exhibit typicality as far as microscopic 
observables are concerned.


Let us finally turn to macroscopic (extensive or intensive) 
observables $A(\phi)$, for instance the total kinetic 
energy of all particles.
Then it is generally expected 
and even provable in some cases \cite{nes25a} 
that the fluctuations 
in (\ref{5}) become negligible for sufficiently large systems 
on the characteristic 
scale of the considered observable
\cite{f2}.
In turn, this means that the difference $A(\phi)-A_{\rm mc}$
on the right-hand side of (\ref{5}) must be negligibly
small for most members $\phi$ of the energy shell, i.e.,
we recover the hallmark of typicality \cite{f3}.

In conclusion, in classical systems typicality is generally
expected to be absent for microscopic and present for
macroscopic observables.
In both cases the essential underlying physics
 -- thermal fluctuations are generically 
 negligible on macroscopic but not on microscopic scales --
appear to be physically very convincing,
but to the best of our knowledge not yet rigorously 
provable in satisfying generality.
In some exceptional cases, they actually may even be wrong, 
as exemplified by the critical (macroscopic) fluctuations 
of systems at a critical point.
From this viewpoint, the problem 
seems in fact very closely related to the derivation of 
general rigorous results regarding the occurrence (or not) of 
phase transitions.

We finally remark that our present results for macroscopic observables
in classical many-body systems 
exhibit some similarities to those by Khinchin in Ref.~\cite{khi60}. 
The main difference is that in Khinchin's approach only non-interacting systems are admitted,
meaning that the 
Hamiltonian
is required to be of the 
very special structure 
$H=\sum_n H_n$, 
where each $H_n$ describes a subsystem (for instance one particle) 
which does not interact with all other subsystems.


%
\vspace*{0.2cm}
\section{Discussion}
\label{s4}

As seen in the first  part of Sec. \ref{s2},
the basic phenomenon of quantum typicality 
arises in any sufficiently high-dimensional subspace 
$\hr_{\! D}$,
independently of any further details of the considered 
quantum system.
A particularly simple example is a harmonic oscillator in 
one dimension (albeit the physical relevance of typicality 
is not obvious in examples like this).
In fact, it is not even necessary that $\hr_{\! D}$
arises in the context of some quantum 
mechanical model system in the first place.
It therefore appears questionable whether 
quantum typicality may generally be understood 
as a consequence of quantum mechanical
entanglement effects, as it is sometimes 
suggested.
Rather, elementary geometrical properties of high 
dimensional Hilbert spaces may be at the root.
If so, there remains no obvious {\em a priori} 
reasons why similar typicality phenomena 
may not arise in classical systems as well.
This question was at the focus of Sec. \ref{s3},
with the main result that classical typicality is generally 
expected to be encountered for macroscopic 
but not for microscopic observables.

A first important implication of the different typicality
properties in quantum and classical systems is that pure states
seem to be very different things in quantum 
and classical mechanics.
In particular, most quantum pure states 
cannot be expected to
approach some
classical pure states in the classical limit.
A second main conclusion is that typicality (in the original 
sense which applies to arbitrary observables) seems to be 
a purely quantum mechanical phenomenon.
According to Sec.~\ref{s21},
quantum fluctuations rather than
entanglement effects (see above)
seem 
to be its
basic physical 
origin.




\vspace*{1cm}
\begin{acknowledgments}
This work was supported by the 
Deutsche Forschungsgemeinschaft (DFG, German Research Foundation)
under Grant No. 355031190 within the Research Unit FOR 2692
and under Grant No. 502254252.
\end{acknowledgments}




\begin{thebibliography}{99}
\bibitem{f1}
Given a orthonormalized basis $\{|n\rangle\}_{n=1}^D$ of $\hr_{\!D}$, 
the considered vectors $|\psi\rangle$ can be written as $\sum_{n=1}^D c_n |n\rangle$,
where $c_n=a_n+i b_n$ are complex coefficients with real and imaginary parts 
$a_n$ and $b_n$, respectively, satisfying the normalization condition 
$\sum_{n=1}^D a_n^2+b_n^2=1$.
Sampling 2D-dimensional vectors $(a_1,b_1,...,a_D,b_D)$ 
according to a uniform distribution on the unit sphere in $\RR^{2D}$ 
thus amounts to an explicit recipe of how uniformly distributed,
normalized $|\psi\rangle\in\hr_{\! D}$ may be generated.
\bibitem{f0}
The mathematical derivation of this entirely general and rigorous
result
can be readily inferred for instance from the findings in
Refs. \cite{llo88} or \cite{gem04} 
together with 
Chebyshev's inequality.
\bibitem{f4}
Moreover, the upper bound (\ref{n1}) can be optimized by working with the
restriction (projection) of $A$ to $\hr_{\! D}$ and
adding a suitable constant to $A$ such that the largest and smallest 
eigenvalues are of equal modulus and opposite sign.
\bibitem{f2}
Intuitively, such a behavior may often be understood as a consequence
of the central limit theorem or some generalization thereof.
\bibitem{f3}
A more precise formulation of this conclusion 
can be inferred from Chebyshev's inequality.
\end{thebibliography}
\end{document}